\documentclass{ws-p8-50x6-00}

\newcommand{\subs}[1]{\ensuremath{\scriptstyle \mathit{#1}}}

\renewcommand{\arraystretch}{1.3}
\newcommand{\FCal}[1][]{\ifthenelse{\equal{#1}{1}}{\textsc{FCal1}}%
                         {\ifthenelse{\equal{#1}{2}}{\textsc{FCal2}}%
                           {\ifthenelse{\equal{#1}{3}}{\textsc{FCal3}}%
                                                  {\textsc{FCal}}%
                           }%
                         }%
                       }%

\newcommand{\alphas}{\ensuremath{\alpha_{s}}}

\newcommand{\dzero}{\ensuremath{\mbox{D\O}}}

\newcommand{\mycsav}{\ensuremath{\langle d^{\,2}\sigma/d\et d\eta \rangle}}

\newcommand{\tdcsav}{\ensuremath{\langle d^{\,3}\sigma/d\et d\eta_{1} d\eta_{2} \rangle}}
\newcommand{\x}{\ensuremath{x}}
\newcommand{\y}{\ensuremath{y}}
\newcommand{\z}{\ensuremath{z}}

\newcommand{\et}{\ensuremath{E_{T}}}

\newcommand{\peta}{\ensuremath{\eta}}				
\newcommand{\aeta}{\ensuremath{|\eta|}}				
\newcommand{\ipb}{\ensuremath{\mathrm{pb}^{-1}}}

\newcommand{\met}{\mbox{${\hbox{$E$\kern-0.63em\lower-.18ex\hbox{/}}}_{T}\,$}}
\newcommand{\metvec}{\mbox{${\hbox{$\vec{E}$\kern-0.63em\lower-.18ex\hbox{/}}}_{T}\,$}}
\newcommand{\metx}{\mbox{${\hbox{$E$\kern-0.63em\lower-.18ex\hbox{/}}}_{x}\,$}}
\newcommand{\mety}{\mbox{${\hbox{$E$\kern-0.63em\lower-.18ex\hbox{/}}}_{y}\,$}}

\newcommand{\etaphi}{\ensuremath{\eta-\varphi}}

\newcommand{\HERWIG}{{\sc Herwig}}
\newcommand{\JETRAD}{{\sc Jetrad}}

\newcommand{\Rsep}{\ensuremath{R_{\subs{sep}}}}
\newcommand{\etal}{{\it et al.}}
\newcommand{\ppbar}{\ensuremath{p\overline{p}}}

\begin{document}
\title{Forward Jets, Dijets, and Subjets at the Tevatron}
\author{Levan Babukhadia}
\address{Physics Department, University of Arizona, Tucson, AZ $\mathit{85721}$, USA\\
         \vskip0.1cm
         \textup{on behalf of the CDF and \dzero\ Collaborations}}
\maketitle
\abstracts{ 
Recent new results on the determination of the rapidity 
dependence of the differential inclusive jet cross 
section, \mycsav,
as a function of jet \et\ in \ppbar\ collisions at 
$\sqrt{s}=1800$ GeV, measured by the \dzero\ detector 
at the Tevatron, are presented along with the comparisons 
to theoretical next-to-leading order (NLO) perturbative QCD 
predictions. Triple differential dijet cross sections,
\tdcsav, at $\sqrt{s}=1800$ GeV, as well as the new results 
on jet structure at $\sqrt{s}=1800$ and $630$ GeV, as 
measured by the CDF and \dzero\ detectors, are also 
discussed.}
\section{Introduction}
\noindent
The last decade of the $20$th century in high energy physics 
will undoubtedly be noted for the impressive progress 
made in both theoretical and experimental understanding 
of collimated streams of particles or ``jets'' resulting 
from inelastic hadron collisions.
The Fermilab Tevatron \ppbar\ Collider, having operated at 
center-of-mass (CM) energies of $1800$ GeV and $630$ GeV, is
a prominent arena for experimentally studying hadronic jets.
Theoretically, jet production in \ppbar\ collisions is 
understood within the framework of quantum chromodynamics
(QCD) as a hard scattering of the constituents of protons, 
quarks and gluons (or partons) that, having undergone the 
collision, manifest themselves as jets in the final state.
Studying various jet properties in the two collider 
experiments, CDF and \dzero, therefore provides stringent 
tests of QCD.
In what follows, we discuss recent results from the two 
experiments, spanning topics from the short to long range 
physics of jets.
\section{Tests of pQCD --- Jet Cross Sections}
\noindent
Perturbative QCD (pQCD) theoretical calculations 
of various jet cross sections~\cite{theory} and new, 
accurately determined parton distribution functions
(pdf's)~\cite{pdfs} add particular interest to the 
corresponding experimental cross section measurements 
at the Tevatron.
These measurements test the short range behavior of QCD, 
the structure of proton in terms of pdf's,
and, if it exists, the substructure of quarks and gluons.
The cross section measurements reported here are based on
integrated luminosities of $87$ and $92$ \ipb\ 
collected by the CDF and \dzero\ experiments, respectively, 
during the $1994$--$95$ Tevatron run.

In both experiments, jets are reconstructed using an 
iterative cone algorithm with a fixed cone radius of 
$\mathcal{R}=0.7$ in \etaphi\ space, where pseudorapidity
\peta\ is related to the polar angle (from the beam line) 
$\theta$ via $\peta=\ln[\cot\theta/2]$ and $\varphi$ is the 
azimuthal angle.
The offline data selection procedure eliminates contamination
from background events caused by electrons, photons, noise, or 
cosmic rays.
This is achieved by applying an acceptance cut on 
the \z--coordinate of the interaction vertex, flagging 
events with large missing transverse energy, and  
applying jet quality cuts.
Details of data selection and corrections due to noise
and/or contamination are described elsewhere~\cite{selection}.

Furthermore, the jet energy scale correction 
removes instrumentation effects associated with calorimeter 
response, showering, and noise, as well as the contribution 
from spectator partons.
The energy scale corrects jets from their reconstructed \et\ 
to their ``true'' \et, on average. 
The smearing effects of the finite calorimeter resolutions
on jet cross sections are removed by an unfolding
procedure.
In \dzero, the scale and resolution corrections are 
determined mostly from the data and are applied in two 
separate steps, while in CDF both corrections are done 
in a single step by means of a Monte Carlo (MC) tuned 
to the data.
\subsection{Rapidity Dependence of the Inclusive Jet 
	    Cross Section}
\begin{figure}[!t] \centering
  \begin{picture}(118,66)  

  \put(-2,2.5){\begin{picture}(70,75)
               \epsfxsize=6.7cm
               \epsfbox{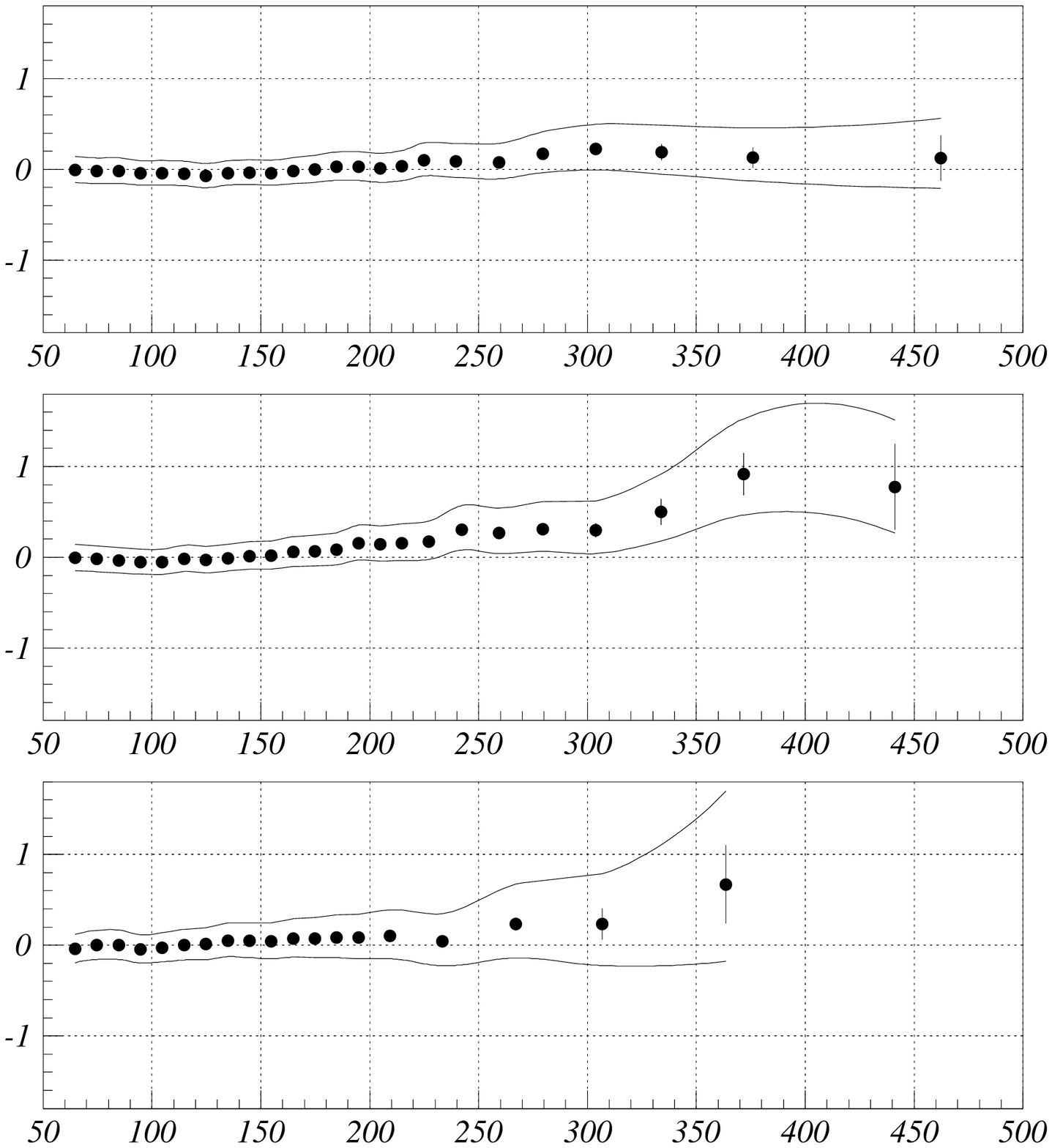}
            \end{picture}}

  \put(55.3,2.5){\begin{picture}(70,75)
               \epsfxsize=6.7cm
               \epsfbox{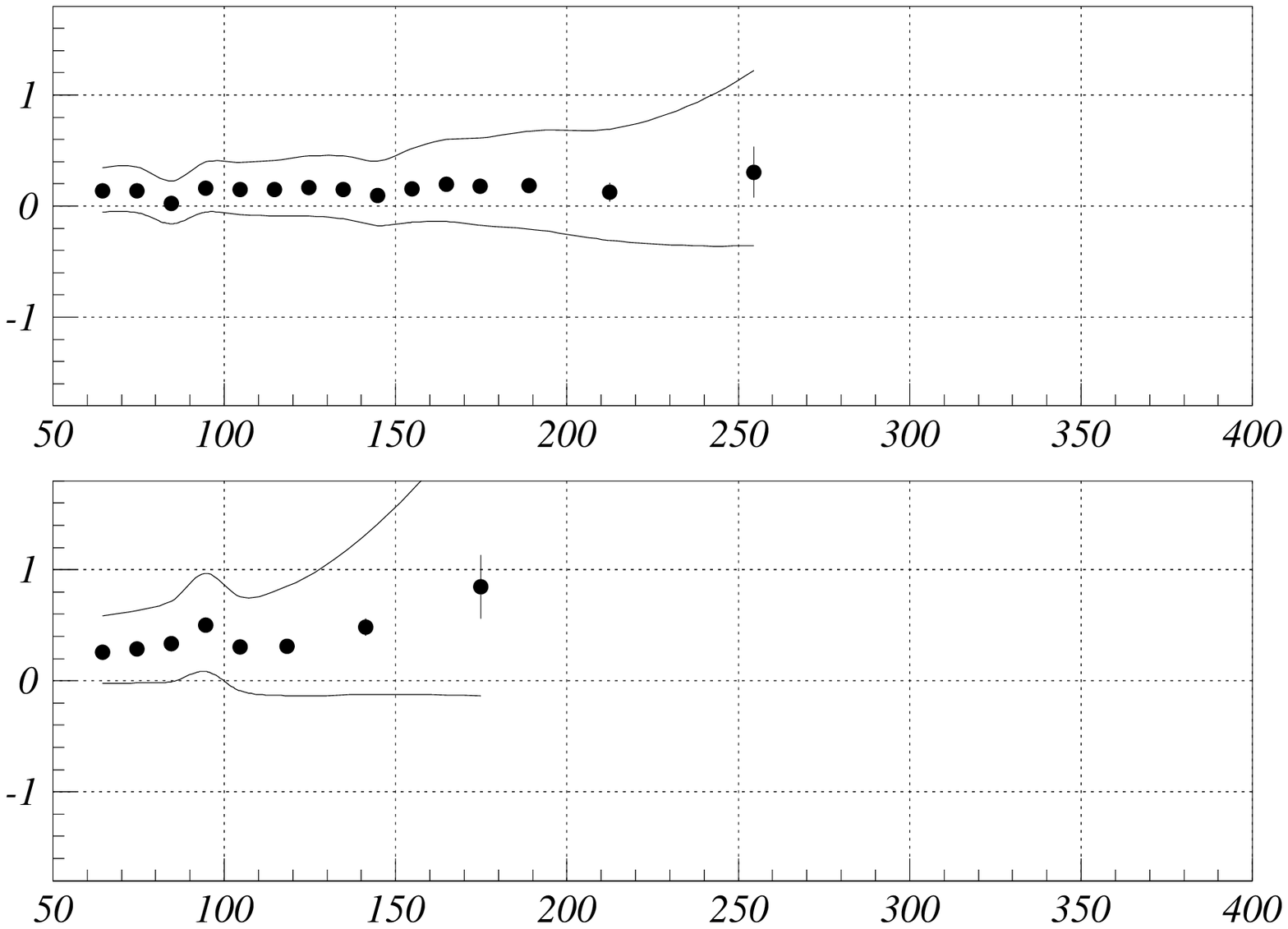}
             \end{picture}}

  \put(7.5,21.5)
    {\makebox(0,0)[l]{\boldmath $\scriptscriptstyle{1.0\,\leq\,\aeta\,<\,1.5}$}}
  \put(113,21.5)
    {\makebox(0,0)[r]{\boldmath $\scriptscriptstyle{2.0\,\leq\,\aeta\,<\,3.0}$}}
  \put(7.5,42.5)
    {\makebox(0,0)[l]{\boldmath $\scriptscriptstyle{0.5\,\leq\,\aeta\,<\,1.0}$}}
  \put(113,42.5)
    {\makebox(0,0)[r]{\boldmath $\scriptscriptstyle{1.5\,\leq\,\aeta\,<\,2.0}$}}
  \put(7.5,63.8)
    {\makebox(0,0)[l]{\boldmath $\scriptscriptstyle{0.0\,\leq\,\aeta\,<\,0.5}$}}

  \put(0,17.5)
    {\rotatebox{90}{\bf {\footnotesize (Data - Theory)/Theory}}}
  \put(46,0.5)
    {\makebox(0,0)[lb]
      {\boldmath $\scriptstyle E_{T} \ \mathrm{(GeV)}$}}
  \put(118,0.5)
    {\makebox(0,0)[rb]
      {\boldmath $\scriptstyle E_{T} \ \mathrm{(GeV)}$}}

  \renewcommand{\arraystretch}{1.0}
  \put(90,66.5)
    {\makebox(0,0)[ct]
      {\begin{tabular}{c}
         {\bf \dzero\ Preliminary} \\ 
         {\bf Run~1B} {\boldmath$\int \! \mathcal{L} dt = 92 \, \ipb$} \\
         Comparisons to {\sc Jetrad} \\
         $\mu_{R}=\mu_{F}=\et^{max}/2$ and \\
         $\Rsep=1.3$.
       \end{tabular} }}
  \renewcommand{\arraystretch}{1.3}
  
  \end{picture}
  \vskip-0.2cm
  \caption{The comparisons between the \dzero\ single inclusive 
	   jet production cross section, \mycsav, as a function of jet \et\ 
	   in five jet \peta\ regions (up to $\aeta=3.0$) and the 
	   $\alphas^{3}$ QCD predictions calculated by \JETRAD\ 
	   with the {\sc CTEQ3M} pdf.}
  \label{fig:rapdep}
  \vspace{-0.5cm}
\end{figure}

\noindent
Recently, \dzero\ has made a new measurement of the 
rapidity dependence of the inclusive single jet cross 
section.
The cross section is determined as a function of jet
transverse energy \et\ in five intervals of jet \aeta, 
up to $\aeta=3.0$, significantly extending previously 
available inclusive jet cross section measurements up 
to $\aeta=0.7$.
The single inclusive jet cross section is calculated 
from the number of jets in each $\eta$--\et\ 
bin, scaled by the 
integrated luminosity, data selection inefficiencies, 
and the unfolding correction.
The experimental measurement in each of the five \aeta\ 
regions is compared to the $\alphas^{3}$ theoretical
predictions by \JETRAD\ (Giele, \etal~\cite{theory}) 
with renormalization and 
factorization scales set equal to each other and
to $\et^{max}/2$ and the parton clustering 
parameter $\Rsep=1.3$.
Comparisons have been made to all recent pdf's 
of the {\sc CTEQ} and {\sc MRST} families~\cite{mytalk}.
Fig.~\ref{fig:rapdep} shows the comparisons with 
the {\sc CTEQ3M} pdf on a linear scale.
The error bars are statistical uncertainties, while the 
error bands indicate $1\sigma$ systematic uncertainties.
Theoretical uncertainties are on the order of 
the systematic errors.
The jet cross cross section, which spans seven
orders of magnitude from the lowest to the 
highest \et's considered, agrees within errors with 
the theoretical pQCD predictions over the full 
dynamical range.
\subsection{Triple Differential Dijet Cross Section}
\noindent
CDF and \dzero\ have measured the triple differential 
dijet cross sections, which are generally less sensitive 
to the pQCD matrix elements and more sensitive to the 
pdf's.
\begin{figure}[p] \centering
  \begin{picture}(118,167)  

  \put(0,93){\begin{picture}(70,75)
               \epsfxsize=6.9cm
               \epsfbox{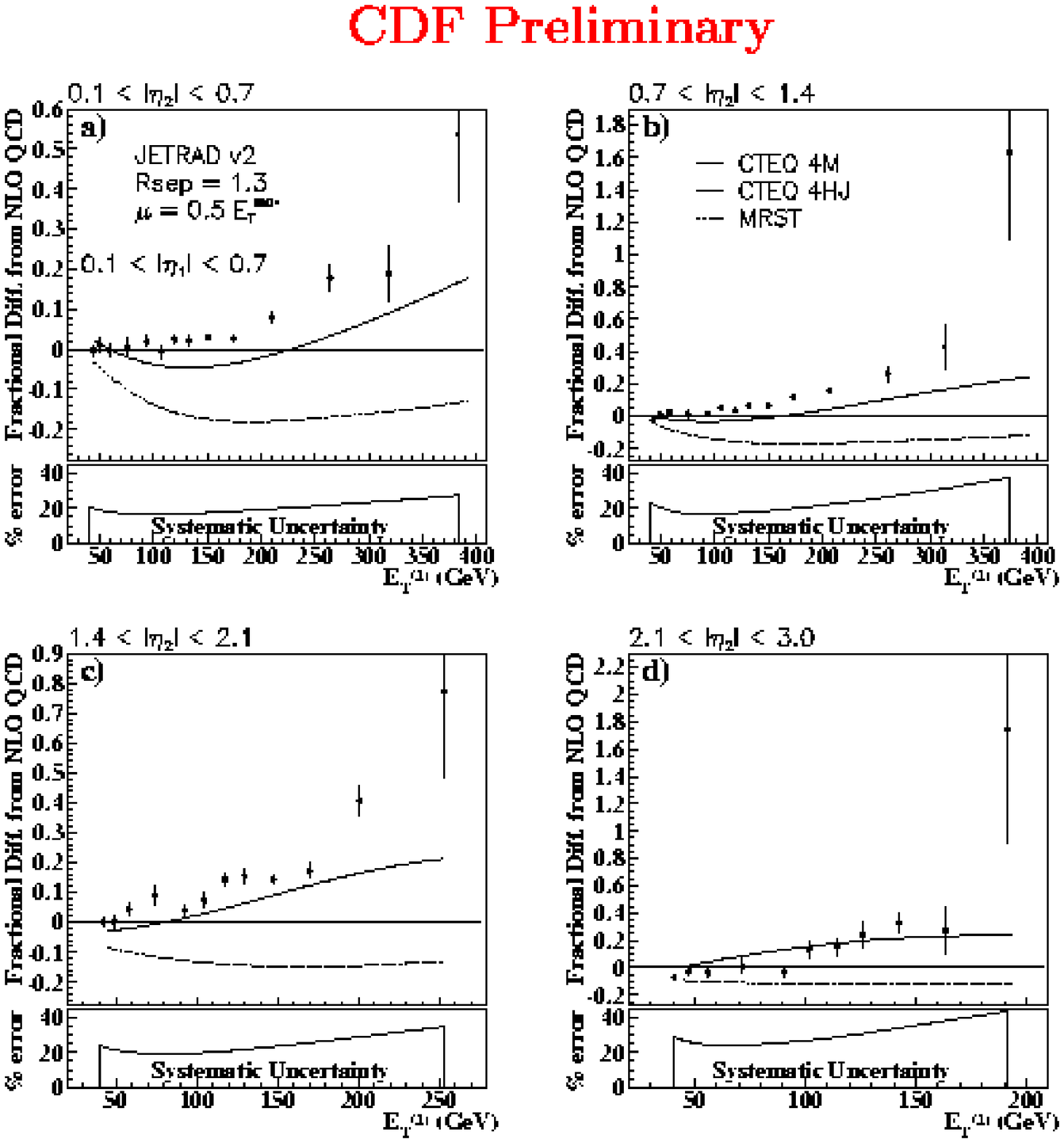}
            \end{picture}}

  \put(74,93)
    {\makebox(0,0)[lb]
      {\begin{minipage}[b]{4.4cm}                      
         \refstepcounter{figure}
         \footnotesize 
         Figure \thefigure.{ Comparisons of the
	 CDF triple differential dijet cross sections
	 to $\alphas^{3}$ QCD \JETRAD\ theoretical 
         calculations with various pdf's, normalized
         to the predictions with {\sc CTEQ4M} pdf.
         Note different \y--scales of
	 the plots corresponding to different \aeta\
         regions ($\aeta<3.0$). Systematic errors are 
	 shown in percent at the bottom of each plot.}
         \label{fig:3dcdf}
       \end{minipage}}}

  \put(-7.5,13){\begin{picture}(70,75)
               \epsfxsize=7.2cm
               \epsfbox{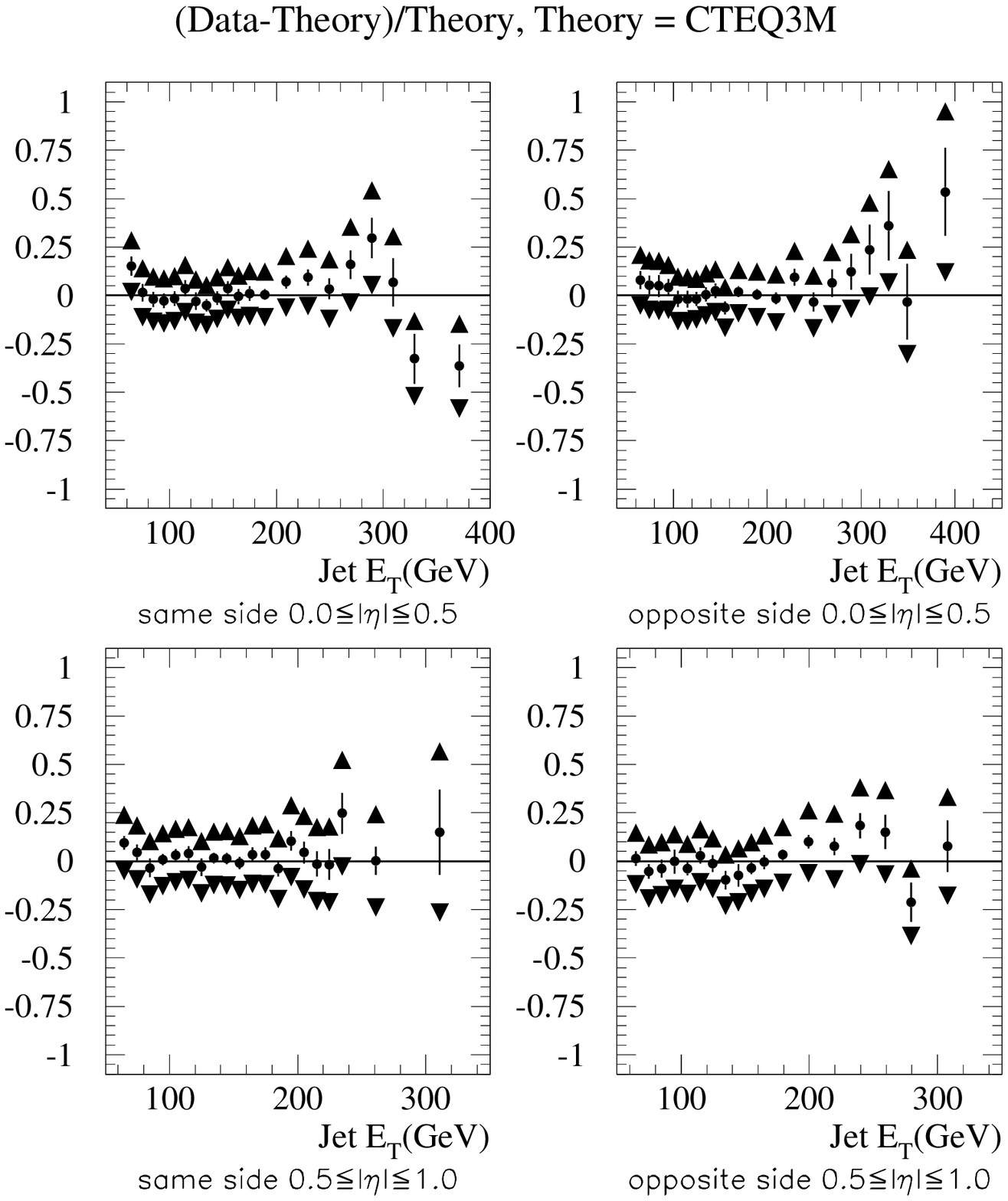}
            \end{picture}}

  \put(52.5,13){\begin{picture}(70,75)
               \epsfxsize=7.2cm
               \epsfbox{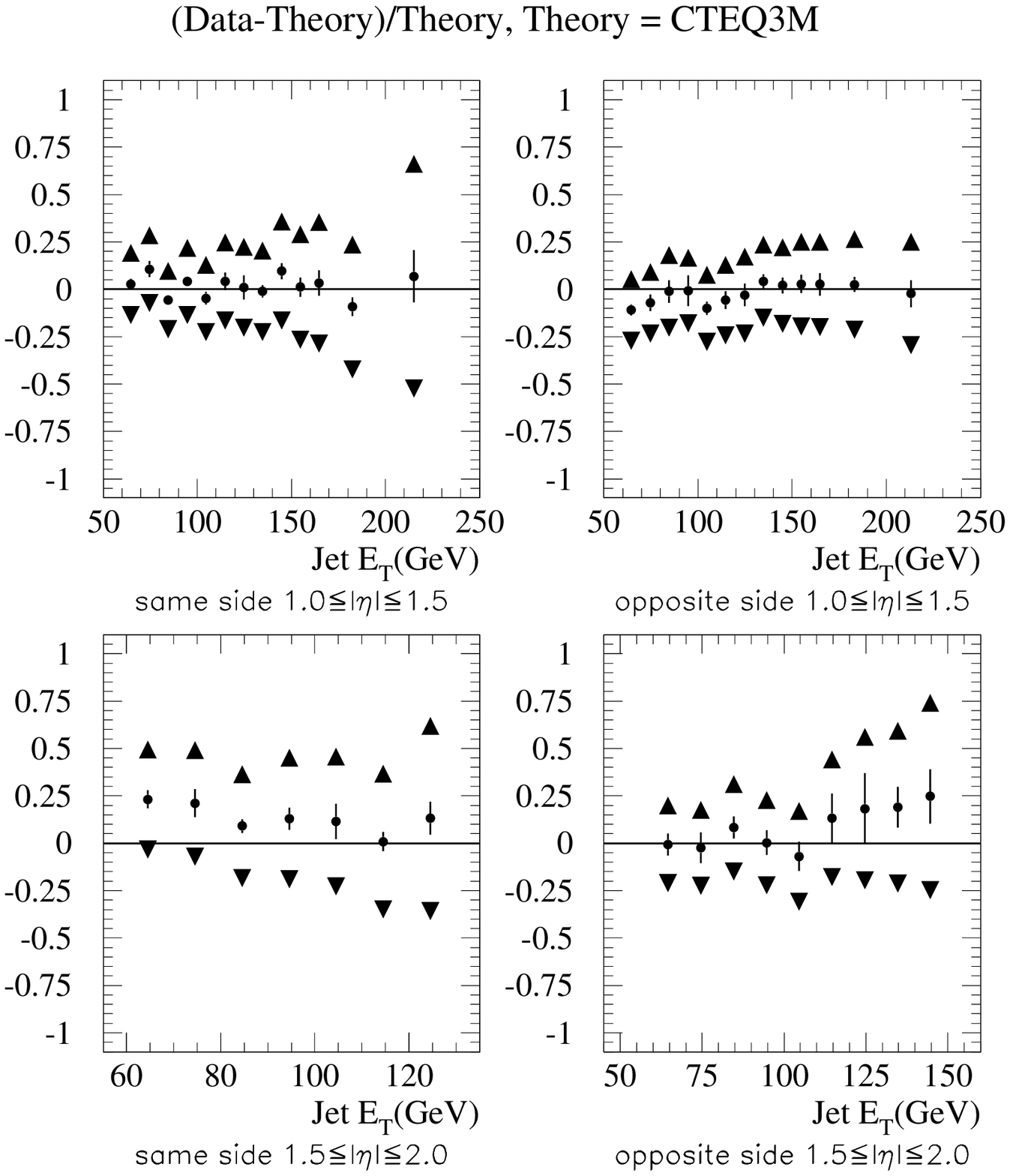}
             \end{picture}}

  \end{picture}
 
  \vskip-1.4cm
  \caption{The comparison between the \dzero\ dijet triple 
	   differential jet cross section as a function of 
	   jet \et\ in four jet \peta\ regions (up to 
	   $\aeta=2.0$) for the SS and OS topologies, and the 
	   $\alphas^{3}$ QCD predictions calculated by 
	   \JETRAD\ with the {\sc CTEQ3M} pdf.}
  \label{fig:3dd0}
\end{figure}
In CDF, the trigger jet has $\et>40$ GeV and is
central ($\aeta<0.7$), while the probe 
jet has $\et>10$ GeV and sweeps all four \aeta\ 
regions considered (up to $\aeta=3.0$).
Fig.~\ref{fig:3dcdf} shows the comparisons of the 
CDF measurement to the $\alphas^{3}$ pQCD predictions
obtained by \JETRAD\ with three different pdf's and 
input parameter values as indicated.
The systematic uncertainties are highly correlated.
The data tend to be higher than pQCD at large \et's.
{\sc CTEQ4HJ} pdf shows better agreement with the data
at high \et's.
Theoretical uncertainties are about as large as the
systematics errors.

In \dzero, the \et's of \emph{both} jets are measured,
even at highest \peta's, so as to be able to also reconstruct
Bjorken \x\ values for the hard scattered partons.
The data are divided into four \aeta\ regions which
are further split into 
Same Side (SS), $\eta_{1}\!\!\cdot\!\eta_{2}>0$, and 
Opposite Side (OS), $\eta_{1}\!\!\cdot\!\eta_{2}<0$, 
dijet events and thus a total of eight cross sections 
are measured (up to $\aeta=2.0$).
Since \et's of both jets are determined, there
are two entries per event in each of the eight 
cross sections.
The comparisons with \JETRAD\ predictions are
shown in Fig.~\ref{fig:3dd0}, where statistical
errors are indicated by the error bars, and the 
upper and lower sets of symbols define the $1\sigma$
systematic error band.
\begin{figure}[!th] \centering
  \begin{picture}(118,60)  

  \put(4,0){\begin{picture}(70,75)
               \epsfxsize=5.9cm
	       \epsfysize=6cm
               \epsfbox{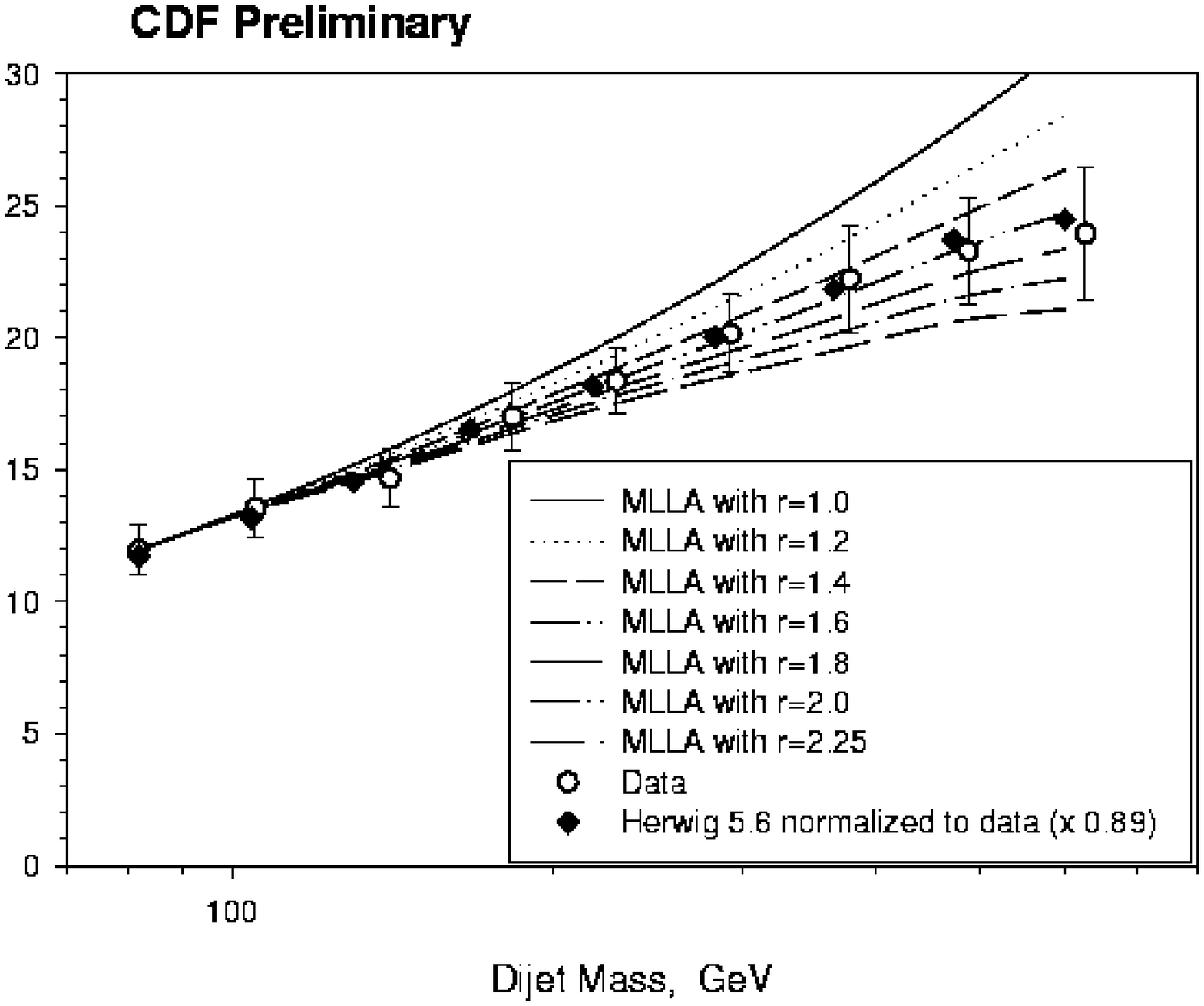}
            \end{picture}}

  \put(69,0)
    {\makebox(0,0)[lb]
      {\begin{minipage}[b]{4.9cm}                      
         \refstepcounter{figure}  
   	 \footnotesize 
         Figure \thefigure.{ Charged multiplicity within
	 a fixed opening angle of $0.466$ rad around the 
	 jet axis in dijets. The data are compared to the
	 \HERWIG\ MC as well as to the MLLA/LPHD predictions
	 with $r=1.0$--$2.25$. A model dependent value of
	 $r=1.7\pm0.3$ is extracted.}
         \label{fig:qgcdf}
       \end{minipage}}}

  \put(0,23){\rotatebox{90}{\bf \footnotesize{Multiplicity}}}

  \end{picture}

\end{figure}
\vspace{-0.5cm}
\section{Tests of the Large Scale QCD --- Jet Structure}
\noindent
While jet cross sections are described by the pQCD 
and pdf's, the structure of a jet is determined by the
large scale QCD behavior which requires non-perturbative 
methods for calculations.
In naive QCD, the ratio $r$ of multiplicities in 
gluon 
and quark jets is expected to be proportional to the
ratio of their 
color charges: $r=C_{A}/C_{F}=9/4$.
In the framework of the Modified Leading Logarithmic 
Approximation (MLLA)~\cite{mlla} with the assumption of
Local Parton--Hadron Duality (LPHD)~\cite{lphd},
the ratio $r$ is expected to decrease with the 
inclusion of NLO and next-to-next-to-leading (NNLO) 
order terms.
CDF has measured the charged track multiplicity in 
dijets
presented in Fig.~\ref{fig:qgcdf} as a function of 
the dijet mass along with the \HERWIG\ MC and 
MLLA/LPHD calculations, the latter with various 
values of the input parameter $r$.
The best fit to the MLLA/LPHD predictions yields 
a model dependent value of $r=1.7\pm0.3$.
CDF has also measured inclusive momentum distributions
of charged particles in dijets and thereby has obtained 
the MLLA/LPHD free parameter 
$Q_{\mathrm{eff}} 
\left( \sim \Lambda_{\mathrm{QCD}} \right) = 
240 \pm 40$ 
MeV.

\begin{figure}[!t] \centering
  \begin{picture}(118,61.5)  

  \put(0,-2){\begin{picture}(70,75)
               \epsfxsize=7cm
               \epsfbox{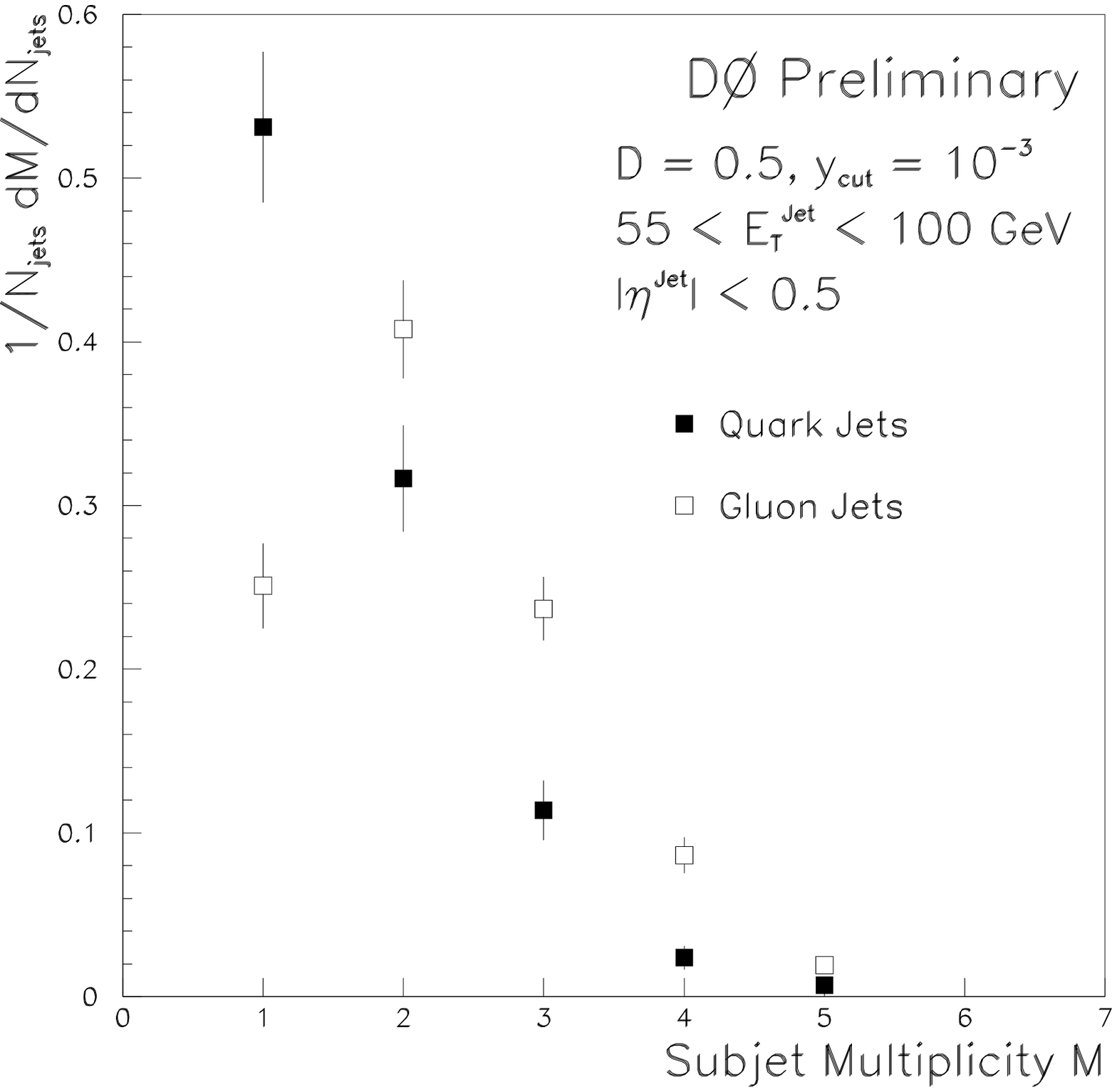}
            \end{picture}}

  \put(69,0)
    {\makebox(0,0)[lb]
      {\begin{minipage}[b]{4.9cm}                      
         \refstepcounter{figure} 
  	 \footnotesize 
         Figure \thefigure.{ Subjet multiplicities in 
	 quark and gluon jets extracted from \dzero\
	 data indicating that, on average, gluon jets 
	 have more subjet clusters than quark jets.
         The $k_{T}$ algorithm is used for jet
         and subjet reconstruction with a ``resolution''
         parameter $y_{\mathrm{cut}}=10^{-3}$ and the
         ``distance'' scale parameter of the jet
         algorithm in \etaphi\ space of $\mathrm{D}=0.5$.}
         \label{fig:qgd0}
       \end{minipage}}}

  \end{picture}
\end{figure}

\dzero\ has measured the gluon and quark subjet
multiplicities ($M_{g}$ and $M_{q}$, respectively) by 
means of a successive combination $k_{T}$ jet 
algorithm~\cite{kt}.
In order to not bias the measurement by gluon/quark 
jet tagging criteria, similar samples are compared \
at $\sqrt{s}=1800$ and $630$ GeV in the same jet \et\ 
($55$--$100$ GeV) and \peta\ intervals ($\aeta<0.5$).
The {\sc Herwig} MC predictions of the fraction of
the gluon jets at the two CM energies of $59$ and
$33\%$, respectively, are used to extract the subjet 
multiplicities in gluon and quark jets shown in Fig.~\ref{fig:qgd0}.
Furthermore, the ratio of gluon to quark subjet 
multiplicities is measured at
$r\equiv(\langle M_{g} \rangle -1)/(\langle M_{q} \rangle -1)=
1.91\pm0.04(\mathrm{stat})^{+0.23}_{-0.19}(\mathrm{sys})$.

\vskip0.5cm
\noindent
{\bf In conclusion,}
CDF and \dzero\ experiments have made several measurements
to test the short as well as the long range behavior of QCD.
Measurements are in good agreement with theoretical
predictions.

\end{document}